\documentclass[manuscript]{aastex}

\shorttitle{Asteroseismic detection of binaries}
\shortauthors{Miglio et al.}

\begin{document}

\title{Prospects for detecting \emph{asteroseismic binaries} in \emph{Kepler} data}

\author{A. Miglio\altaffilmark{1}, W. J. Chaplin\altaffilmark{1}}

\affil{School of Physics and Astronomy, University of Birmingham,
  Edgbaston, Birmingham, B15 2TT, UK}

\author{R. Farmer, U. Kolb}

\affil{Department of Physical Sciences, The Open University,
  Walton Hall, Milton Keynes MK7 6AA, UK}

\author{L. Girardi}

\affil{INAF-Osservatorio Astronomico di Padova, Vicolo
  dell'Osservatorio 5, I-35122 Padova, Italy}

\author{Y. Elsworth\altaffilmark{1}}

\affil{School of Physics and Astronomy, University of Birmingham,
  Edgbaston, Birmingham, B15 2TT, UK}

\author{T. Appourchaux}

\affil{Institut d'Astrophysique Spatiale, UMR8617, Universit\'e Paris
  XI, B\^atiment 121, 91405 Orsay Cedex, France}

\author{R. Handberg\altaffilmark{1}}

\affil{School of Physics and Astronomy, University of Birmingham,
  Edgbaston, Birmingham, B15 2TT, UK}

\altaffiltext{1}{Stellar
  Astrophysics Centre (SAC), Department of Physics and Astronomy,
  Aarhus University, Ny Munkegade 120, DK-8000 Aarhus C,
  Denmark}

\begin{abstract}

{Asteroseismology may in principle be used to detect unresolved
  stellar binary systems comprised of solar-type stars and/or red
  giants. This novel method relies on the detection of the presence of
  two solar-like oscillation spectra in the frequency spectrum of a
  single lightcurve. Here, we make predictions of the numbers of
  systems that may be detectable in data already collected by the NASA
  \emph{Kepler} Mission.} Our predictions, which are based upon
TRILEGAL and BiSEPS simulations of the \emph{Kepler} field of view,
indicate that as many as 200 or more ``asteroseismic binaries'' may be
detectable in this manner. Most of these binaries should be comprised
of two He-core-burning red giants. Owing largely to the limited
numbers of targets with the requisite short-cadence \emph{Kepler}
data, we expect only a small number of detected binaries containing
solar-type stars.  The predicted yield of detections is sensitive to
the assumed initial mass ratio distribution of the binary components
and therefore represents a sensitive calibration of the much debated
initial mass ratio distribution near mass ratio unity.

\end{abstract}

\keywords{binaries: general --- Galaxy: stellar content --- stars:
  evolution --- stars: statistics --- surveys}

\section{Introduction}
\label{sec:intro}

The NASA \emph{Kepler} Mission has provided a wealth of data for
studying stellar binary systems. More than 2000 eclipsing binaries
have so far been found \citep{Prsa11, Slawson11}. Circumbinary planets
have also been discovered in several of these systems (e.g., see
\citealt{Welsh12, Orosz12}). The \emph{Kepler} eclipsing binary
catalogue\footnote{see {\tt http://keplerebs.villanova.edu/}} provides
information on binaries with periods up to a few hundred days.  In
this paper we discuss the prospects for using asteroseismology as a
novel way of discovering non-eclipsing binary systems in numbers that
would potentially provide a useful statistical sample, out to periods
much longer than the eclipsing binary catalogue.

Asteroseismology has been one of the major successes of
\emph{Kepler}. It has in particular revolutionized the study of stars
that show solar-like oscillations, with detections made in several
hundred cool main-sequence and subgiant stars \citep{Chaplin11a,
  Chaplin14}, and over ten-thousand red giants (e.g., see
\citealt{Bedding11, Hekker11, Huber11, Huber13a}). These stars present
rich spectra of overtones in the frequency-power spectrum.

\begin{figure}
\includegraphics[width=.8\textwidth]{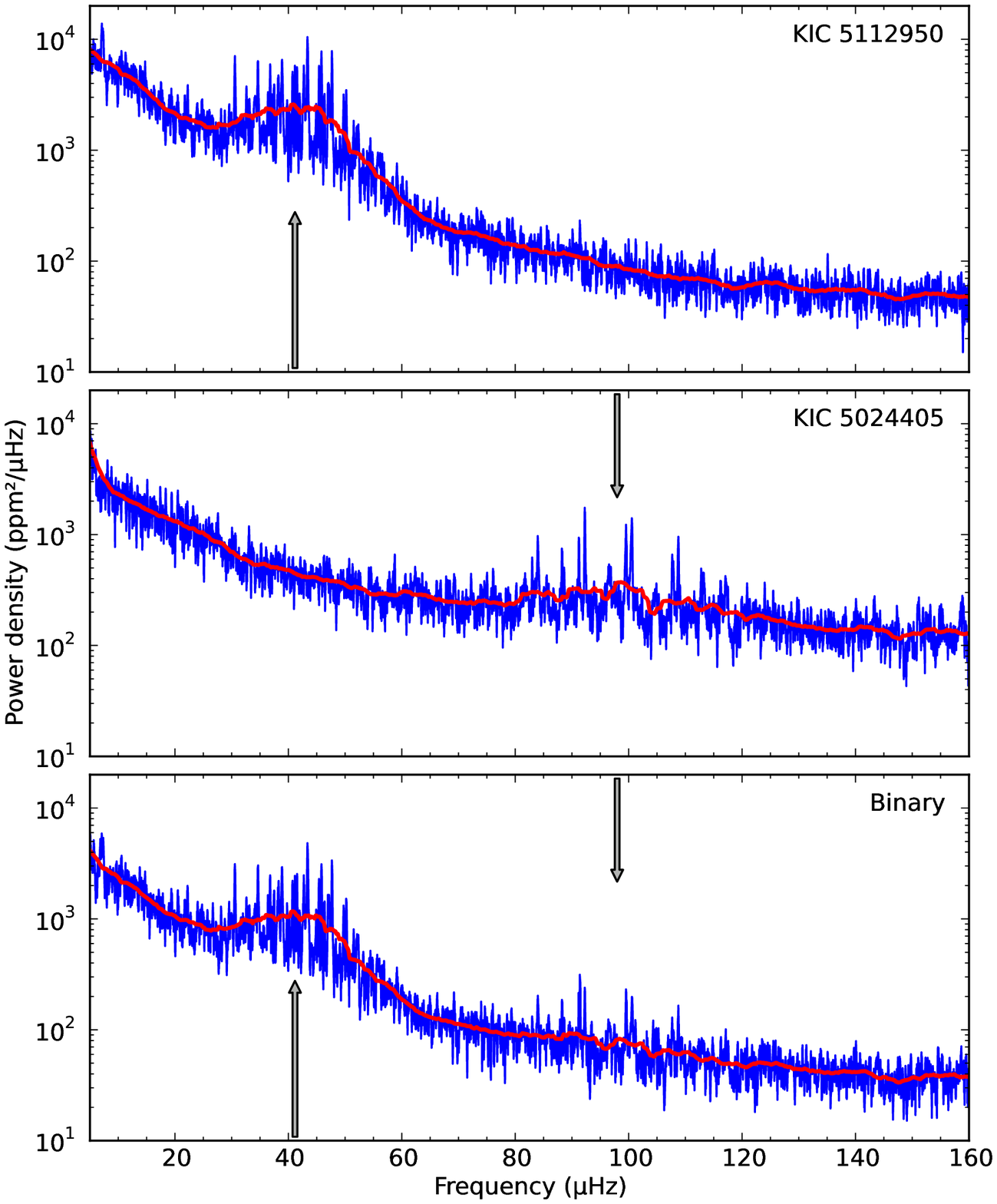}

\caption{Illustration of the concept: artificial asteroseismic binary
  created by using \emph{Kepler} data on two red giants with detected
  oscillations in the open cluster NGC\,6819, KIC\,5024405 and
  KIC\,5112950 (from Fig.~2 of \citealt{Stello10}). Bottom panel:
  frequency-power spectrum of the combined lightcurve. Upper two
  panels: spectra of the original two lightcurves. Arrows mark the
  locations of the solar-like oscillations for each star.}

\label{fig:combined}
\end{figure}

Asteroseismology may be used to discover candidate binaries by
detecting the presence of two solar-like oscillation spectra in the
frequency spectrum of a single \emph{Kepler} lightcurve. The
oscillation spectra may be well separated in frequency; or they may
overlap, with the individual oscillation peaks of the two stars
interspersed. {To illustrate the concept, we have created an
  artificial example by using \emph{Kepler} data on two red giants in
  the open cluster NGC\,6819. We combined the lightcurves of two
  cluster members -- KIC\,5024405 and KIC\,5112950 -- with detected
  oscillations (from Fig.~2 of~\citealt{Stello10}).  The bottom of our
  Fig.~\ref{fig:combined} shows the frequency-power spectrum of the
  combined lightcurve (i.e., the simulated asteroseismic binary),
  whilst the upper two panels show the spectra of the two original
  lightcurves. Arrows mark the locations of the solar-like
  oscillations.}  The analysis required to discriminate the
constituent oscillation frequencies of stars comprising asteroseismic
binaries is beyond the intended scope of the present paper, although
we add that in some scenarios -- when the oscillations spectra are
well separated in frequency -- it is a fairly trivial
matter. {That would have been the case for our artificial
  example: the oscillations are clearly separated (because
  KIC\,5112950 is about 2.5-times more luminous than KIC\,5024405).}

Our aim in this paper is to estimate -- subject to different
assumptions about the underlying binary population -- the number of
binaries in the \emph{Kepler} field where both components would be
expected to show detectable solar-like oscillations in \emph{Kepler}
data. Our results are based on predictions from two stellar population
simulations of the \emph{Kepler} field, to which we applied target
selection criteria based on those used to select the real
\emph{Kepler} target list. We consider binaries comprised of
solar-type stars and/or red giants. Imposing the seismic detectability
constraint for both components strongly favours binaries with mass
ratio near unity. The predicted number therefore represents a very
sensitive calibration of the much debated initial mass ratio
distribution (IMRD; e.g. \citet{goodwin13} and references therein)
near mass ratio unity.

The layout of our paper is as follows. In Section~\ref{sec:popsyn} we
discuss the population synthesis predictions made by TRILEGAL
\citep{Girardi05, Girardi12} (Section~\ref{sec:tri}) and BiSEPS
\citep{Farmer13} (Section~\ref{sec:biseps}); and target selection
applied to the raw predictions (Section~\ref{sec:targ}). In
Section~\ref{sec:det} we explain how we used the data on the intrinsic
properties of the synthetic target-selected populations to make the
asteroseismic detection predictions. Section~\ref{sec:res} presents
results on the asteroseismically detected synthetic binaries. We end
in Section~\ref{sec:disc} with concluding remarks on the implications
of our results for analysis and exploitation of the real, available
\emph{Kepler} data.

 \section{Population synthesis predictions}
 \label{sec:popsyn}

 \subsection{TRILEGAL}
 \label{sec:tri}

We used TRILEGAL \citep[see][for a detailed description]{Girardi05,
  Girardi12} to simulate the Milky Way. TRILEGAL comprises a geometric
model for the principal components, e.g., the thin and thick discs,
halo and bulge, with constituent populations having a prescribed star
formation rate and age-metallicity relation. The dimensions of the
components were calibrated from selected real observations (e.g.,
where reddening is small and photometric incompleteness is not an
issue). The implicit assumption is made that, aside from interstellar
dust, the Galactic components are smoothly distributed, and uniform in
their distributions of age and metallicity.

Standard parameters were used for each component of the Milky
Way. Stellar populations were simulated for the fields covered by each
of the 21 five-square-degree \emph{Kepler} CCD pairs.  Interstellar
extinction was modelled at infinity \citep{Schlegel98}, and assumed to
arise from a dust layer having a 100-pc scale height.  Stellar
magnitudes were estimated in the \emph{Kepler} bandpass using the
known instrumental response function. To mimic the selection function
used to compile the \emph{Kepler} target list, the synthetic
population was filtered using the procedure described below in Section
\ref{sec:targ}.

Non-interacting binaries are included in the stellar populations
simulated with TRILEGAL.  For each star, there is a probability
$f_{\rm b}$ that it belongs to a binary system. The secondary star has
the same age and metallicity as the primary, and its mass is drawn
from a flat IMRD in the interval $[b_{\rm b},1]$. Default values of
$f_{\rm b}$ and $b_{\rm b}$ are 0.3 and 0.7, which we also adopt in
our reference simulations.

\subsection{B{\protect i}SEPS}
\label{sec:biseps}

The BiSEPS code \citep[see][for a detailed description]{willems02,
  willems04, willems06, Farmer13} employs the semi-analytical
description for single and binary star evolution by \citet{hurley00}
and \citet{hurley02} to create a large library of evolutionary
tracks. The resulting Galactic stellar population models therefore
include evolving and interacting binaries in a self-consistent way.
The evolutionary tracks are weighted according to initial distribution
functions and convolved with a star formation history.  We adopt
canonical initial distribution functions for (primary) mass and
orbital separation, while the secondary mass is selected from a
power-law IMRD, $\chi(q)=(s+1)q^s$, where $q$ is the initial mass
ratio and $s$ is a free parameter. We consider simulations computed
with $s=-0.5, 0,$ and 1, assuming that 50\% of all systems form as
binaries.

To model the \emph{Kepler} field content we take into account
contributions by the thin disc and thick disc populations, adopting
hydrogen abundance $X=0.70$ for all stars and representative
metallicities $Z=0.02$ and $Z=0.0033$, respectively. The extinction is
calculated following \citet{drimmel03}, with bolometric corrections
and extinction coefficients taken from \citet{girardi08}.

\subsection{Target Selection}
\label{sec:targ}

To model the impact that the \emph{Kepler} target selection procedure
has on the simulated binary sample we tested the synthetic
\emph{Kepler} field against the same criteria that were used to
compile the actual \emph{Kepler} target list from the \emph{Kepler}
Input catalogue (KIC)\footnote{\tt
  http://www.cfa.harvard.edu/kepler/kic/kicindex.html}.  The
rank-order is determined by the minimum detectable planet radius
(neglecting intrinsic stellar noise) for observations lasting the full
duration of the nominal \emph{Kepler} mission.

The selection process involves the creation of synthetic full-frame
images -- which also include detector noise and zodiacal noise -- to
determine the optimum aperture that maximizes the signal-to-noise
ratio for each target. The ranking is based on stellar parameters
estimated from the synthetic colors of each target, following
\emph{Kepler}'s Stellar Classification Program \citep{brown11}.  We
modelled binaries as being comprised of two point sources, with
effective magnitudes calculated from the summed fluxes of the stars in
the considered wavelength band.

The BiSPES model applies a set of colour corrections to its
synthetic stars to force better agreement with the actual KIC in
colour-colour space; this step was not needed for the TRILEGAL model.

Full details of the target selection procedure are presented in
\citet{Farmer13}.

 \section{Predictions of asteroseismic detectability}
 \label{sec:det}

We applied the procedure in \citet{Chaplin11b} to every simulated star
in the target-selected samples to predict whether we would expect to
make an asteroseismic detection. The procedure uses as input the
fundamental properties of each modelled target -- be it a single field
star or both components of a binary -- as well as the simulated
\emph{Kepler} apparent magnitude, and the assumed duration of the
observations, to estimate what would be the observed photometric
signal amplitude due to solar-like oscillations, granulation and shot
noise. From these estimates we may calculate the likelihood of making
a robust detection of solar-like oscillations. As in
\citet{Chaplin11b}, we flag a detection as made when the estimated
probability of detection is higher than 90\%.

We corrected the predicted amplitudes of the oscillations and
granulation signals of targets in binaries to allow for the dilution
of the observed signal due to the presence of the companion
star. These corrections were made in the \emph{Kepler} bandpass, using
the bolometric corrections in \citet{Ballot11}.

We adopted two different assumed observation durations, depending on
the type of target.  The 58.85-sec \emph{Kepler} short-cadence (SC)
data are needed to detect oscillations in cool main-sequence and
subgiant stars, since the dominant oscillations have periods of the
order of minutes. These short periods are not accessible to the
29.4-minute long-cadence (LC) data (for which the Nyquist frequency is
$\simeq 283\,\rm \mu Hz$).  Due to the target-limited nature of the SC
data, around 2000 targets identified in the KIC as solar-type stars
were observed for only one month at a time during the asteroseismic
survey conducted in the first 10\,months of science operations. To
mimic this limitation of the real target sample, we therefore set the
observation time to 1\,month for any simulated star that would require
SC data to detect its oscillations. The threshold for detection in SC
lies at the base of the red-giant branch. We set the observation duration to
4\,years for more evolved targets that would need only the LC data to
detect their oscillations.

 \section{Results}
 \label{sec:res}

\begin{figure*}
\includegraphics[width=.5\textwidth]{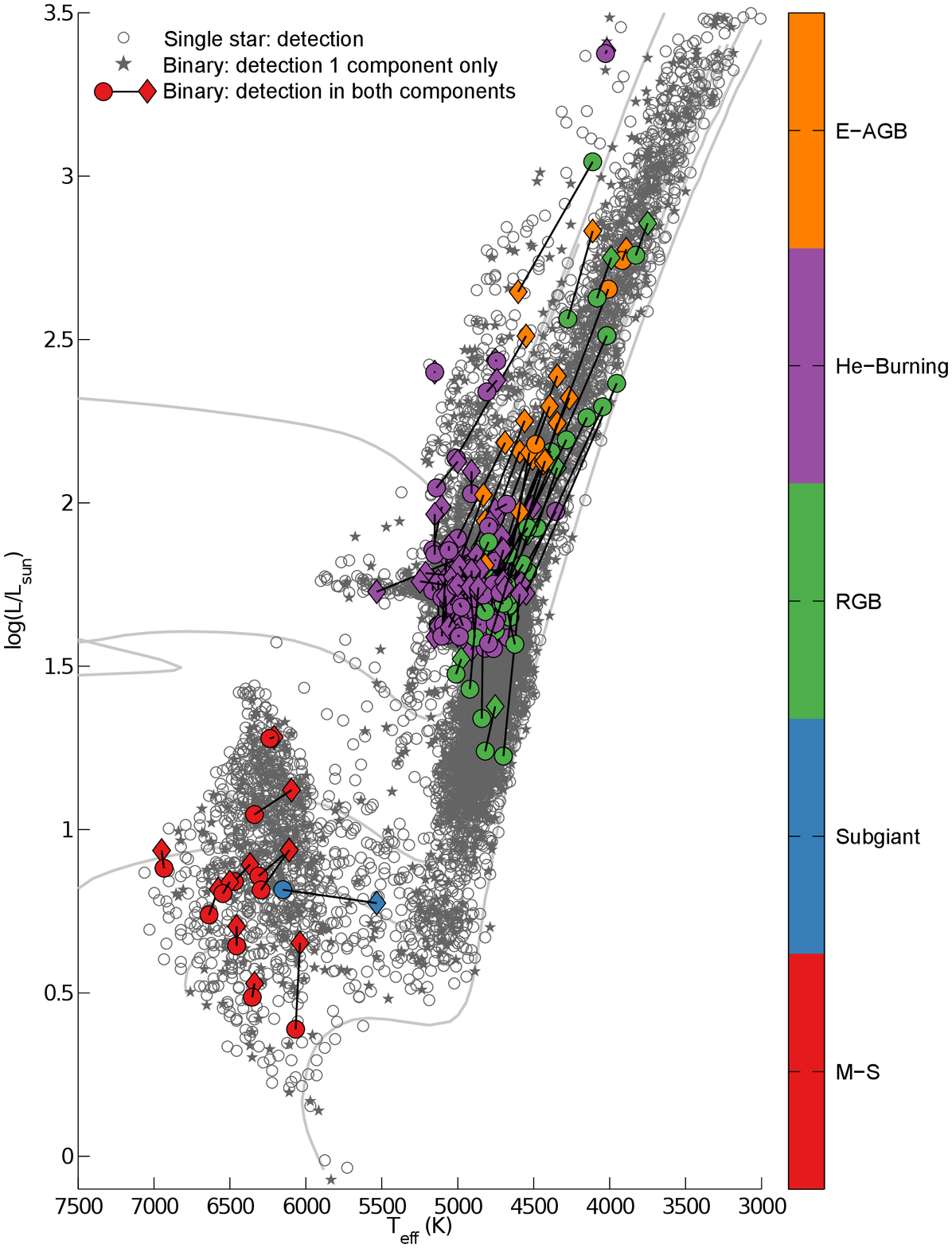}
\includegraphics[width=.5\textwidth]{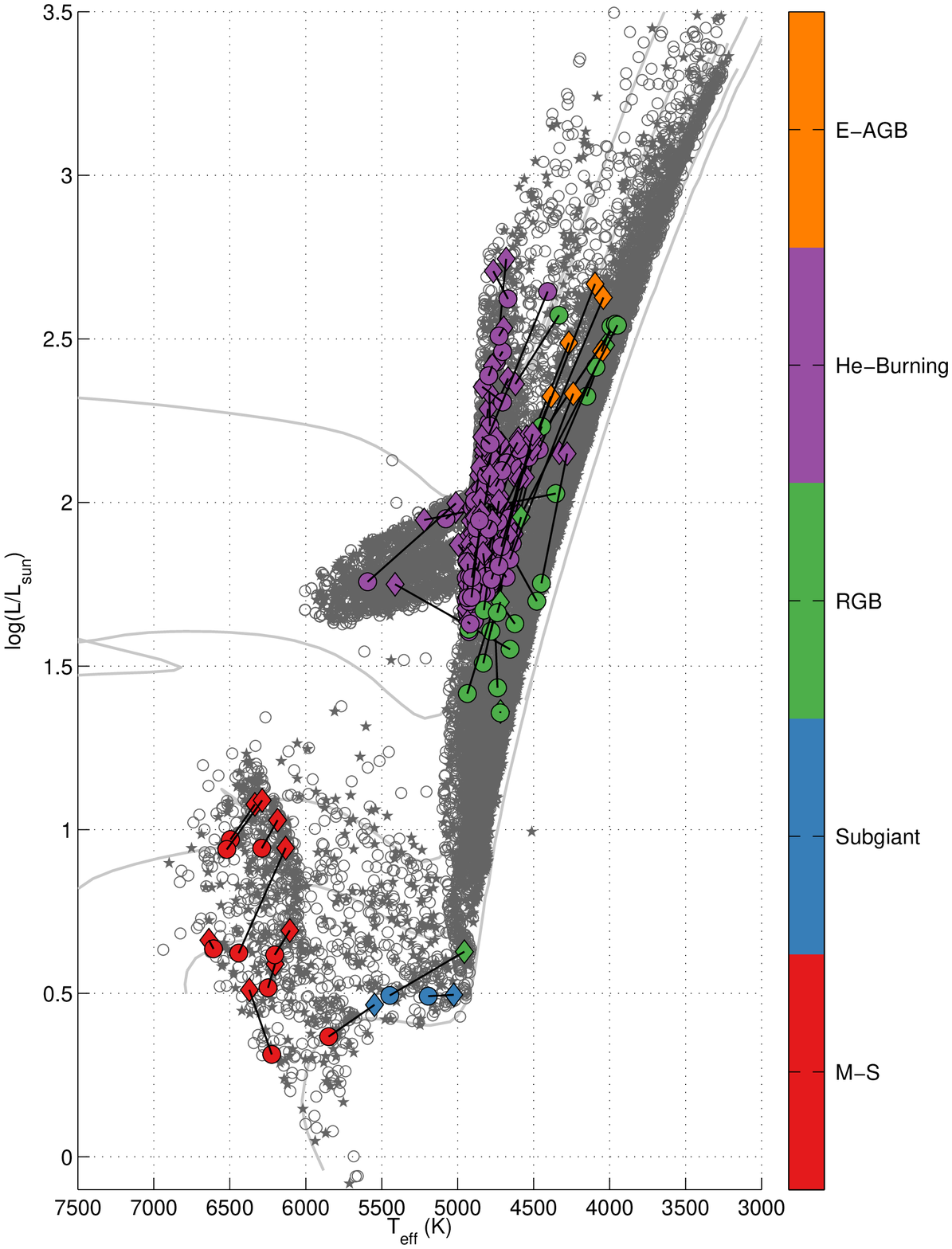}

\caption{Left-hand panel: Hertzsprung-Russell diagram of a synthetic
  population computed with TRILEGAL. Single stars with predicted
  detectable oscillations are shown as open circles. Primary
  components of binaries in which oscillations can only be detected in
  one component are represented by stars.  Couples of filled circles
  and diamonds connected by solid lines indicate members of binary
  systems for which oscillations are expected to be detectable in both
  components, i.e., asteroseismic binaries.  Color illustrates the
  evolutionary state of each component. Right-hand panel: Same as
  left-hand panel, but showing results obtained with B{\protect i}SEPS
  {assuming $s=0$ (see Section~\ref{sec:biseps})}.}

\label{fig:hrd}
\end{figure*}

Fig.~\ref{fig:hrd} is a Hertzsprung-Russell diagram showing members of the
target-selected TRILEGAL and BiSEPS samples with predicted
asteroseismic detections. Results are shown from simulations in which
the binary populations were given a flat IMRD. Grey stars (circles)
mark cases where a detection was predicted for the more luminous
component of a binary (a single field star). The coloured symbols
connected by solid lines mark cases where \textsl{both} components of
a binary showed detectable solar-like oscillations.

\begin{figure}

\includegraphics[angle=0, width=.5\textwidth]{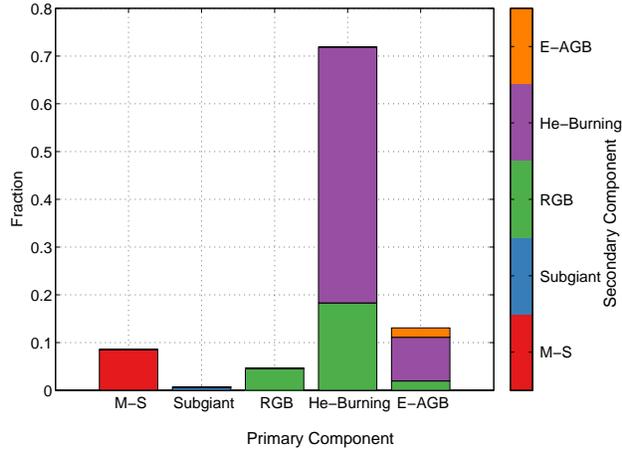}

\caption{Stacked histograms showing the evolutionary state of the
  secondary component of detectable seismic binaries as a function of
  the evolutionary state of the primary.}

\label{fig:histo_ev}
\end{figure}

Results given by the target-selected sample from the TRILEGAL
population, having $f_{\rm b}=0.3$ and $b_{\rm b}=0.7$, predict that
there will be around 200 detectable asteroseismic binaries in the
\emph{Kepler} LC data. This corresponds to just over 1\,\% of the
20,000 red giants flagged as showing a seismic detection. Most of
these asteroseismic binaries will be comprised of two stars in the
He-core burning, or red-clump, phase (see Fig.~\ref{fig:histo_ev}).
Similar results were returned by the BiSEPS population when an IMRD
with exponent $s=1$ was adopted. Absolute numbers drop to around 120
(60) when $s =0$ ($s=-0.5$). The synthetic target-selected samples
include a small fraction of binaries that are either in a phase of
mass transfer or have undergone such a phase in the past. None of
these objects are detectable as asteroseismic binaries.

Only a handful of binaries will be detectable in the main-sequence and
subgiant phase. This is due both to the restrictions on the number of
suitable targets with the requisite SC data (see
Section~\ref{sec:det}), and the lower observable asteroseismic
signal-to-noise ratios in these stars, compared to red giants (which
makes seismic detection of binaries much harder).

We may gain some insights on the likely robustness of the predicted
\textsl{absolute} numbers of binary detections by comparing predicted
numbers of individual detections with those already made from the
\emph{Kepler} data. The simulations predict detections in around 1500
main-sequence and subgiant stars (note we have compensated for the
fact that the simulated target selected samples are around 20\,\%
larger than the real \emph{Kepler} target sample). The predicted
number of detections is approximately a factor of two higher than the
number of recorded detections \citep{Chaplin11a, Verner11,
  Huber13}. Some of this difference may be attributed to the
restricted number of SC target slots that were available to the real
asteroseismic survey; and the selection procedure for the
asteroseismic targets (which we have not reproduced here). There will
also be a contribution due to stellar magnetic activity, which reduces
the detectability of the solar-like oscillations
\citep{Chaplin11c}. This is an effect we have not allowed for in our
simulations.  Our conclusions for the SC data will remain the same in
spite of these differences: we expect very few seismic binary
detections in SC.

The impact of activity will be much less important for the more
evolved cohort of stars observed in LC, where the simulations predict
detections in more than 20,000 giant stars. Detections have already
been recorded in around 16,000 red giants. \citet{Stello13} report on
the asteroseismic analysis of more than 13,000 red giants (see also
\citealt{Hekker11}); whilst another 3000 detections have recently been
made from the analysis of targets that were previously unclassified in
the KIC \citep{Huber13a}. There are undoubtedly other cohorts of
\emph{Kepler} targets, containing red-giant stars, that have not yet
been subjected to asteroseismic analysis (e.g., the so-called ``add
back in'' stars at the bottom of the \emph{Kepler} target list). The
absolute predictions for the LC data are therefore likely to be fairly
robust, and hence the \emph{Kepler} LC data on red giants should
provide testable diagnostics of the underlying binary population,
i.e., by giving detections in numbers that are sufficient for drawing
statistical inference.

A crucial consideration for drawing any inferences from the real data
is of course the likely occurrence rate of false-positives, i.e., the
fraction of seismically detected ``binaries'' we would actually expect
to be comprised of spatially coincident -- at the level of the
4\,arcsec pixel size of \emph{Kepler} -- but physically unassociated,
single stars. The false-positive rate is limited considerably by the
fact that only the 20,000 or so single stars with detectable
oscillations can potentially contribute to it.  We estimate that we
should expect only a few false-positives (the median expectation value
is $\approx 5$) in the LC data. Taking the predictions above, this
translates to a false-positive rate of less than 10\,\% of detectable
asteroseismic binaries, and in some cases significantly lower
depending on the properties of the underlying binary population and
hence absolute numbers of expected detections.  It is also worth
adding that the fraction of resolvable binaries will be close to zero,
i.e., the contribution to false-negatives from resolved binaries is
likely to be negligible.

\begin{figure}
 \includegraphics[angle=0, width=.45\textwidth]{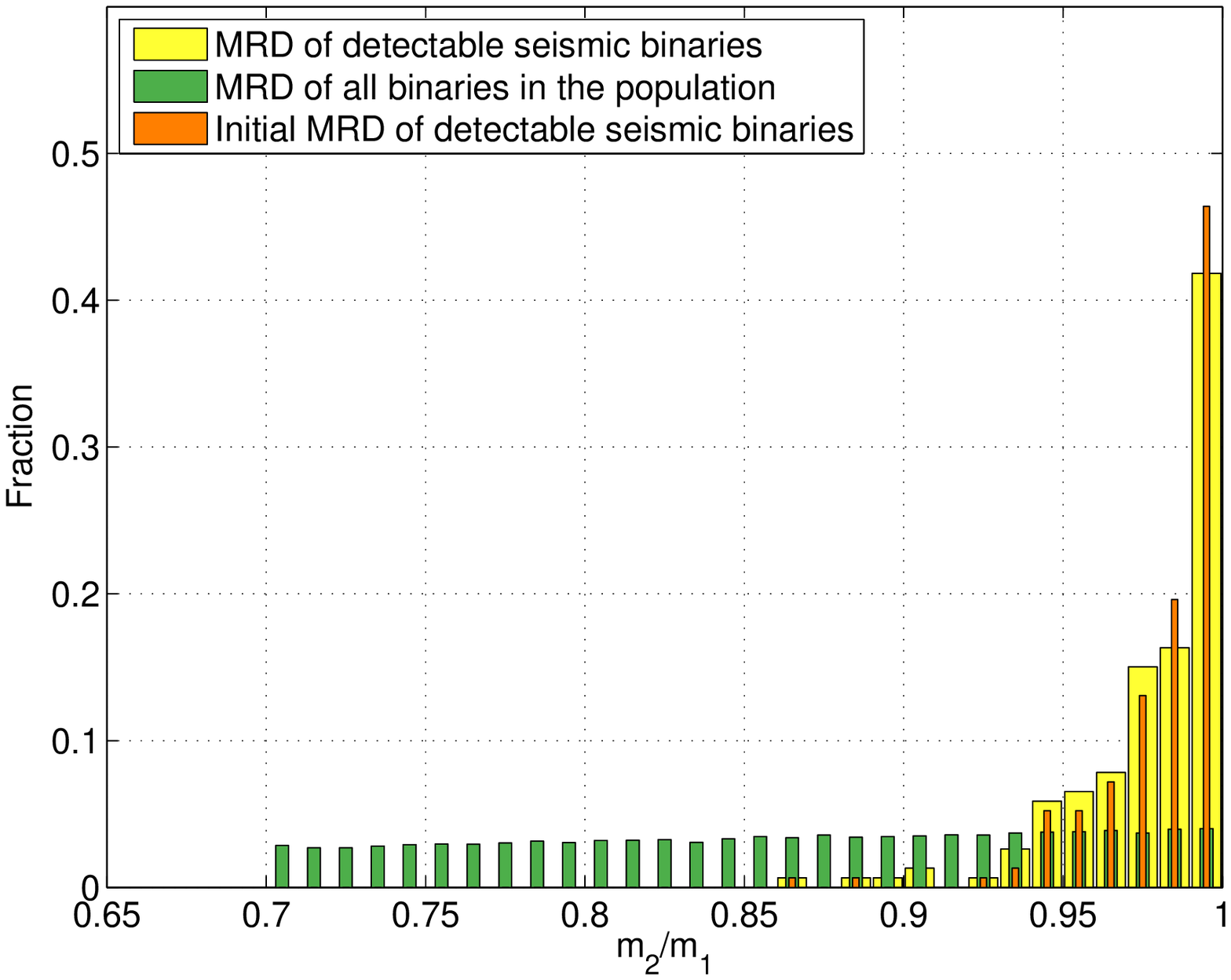}
 \includegraphics[width=.45\textwidth]{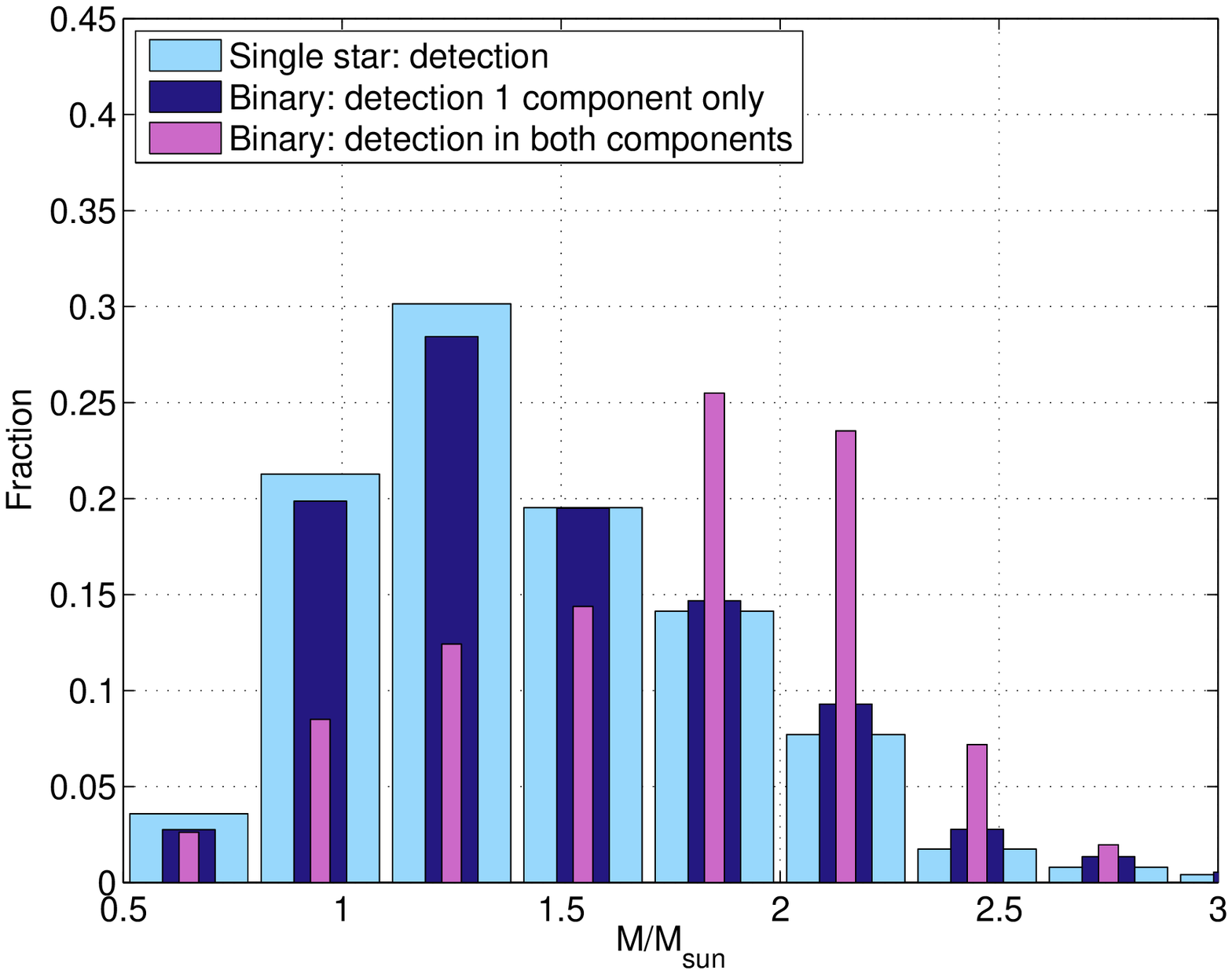}
 \includegraphics[width=.45\textwidth]{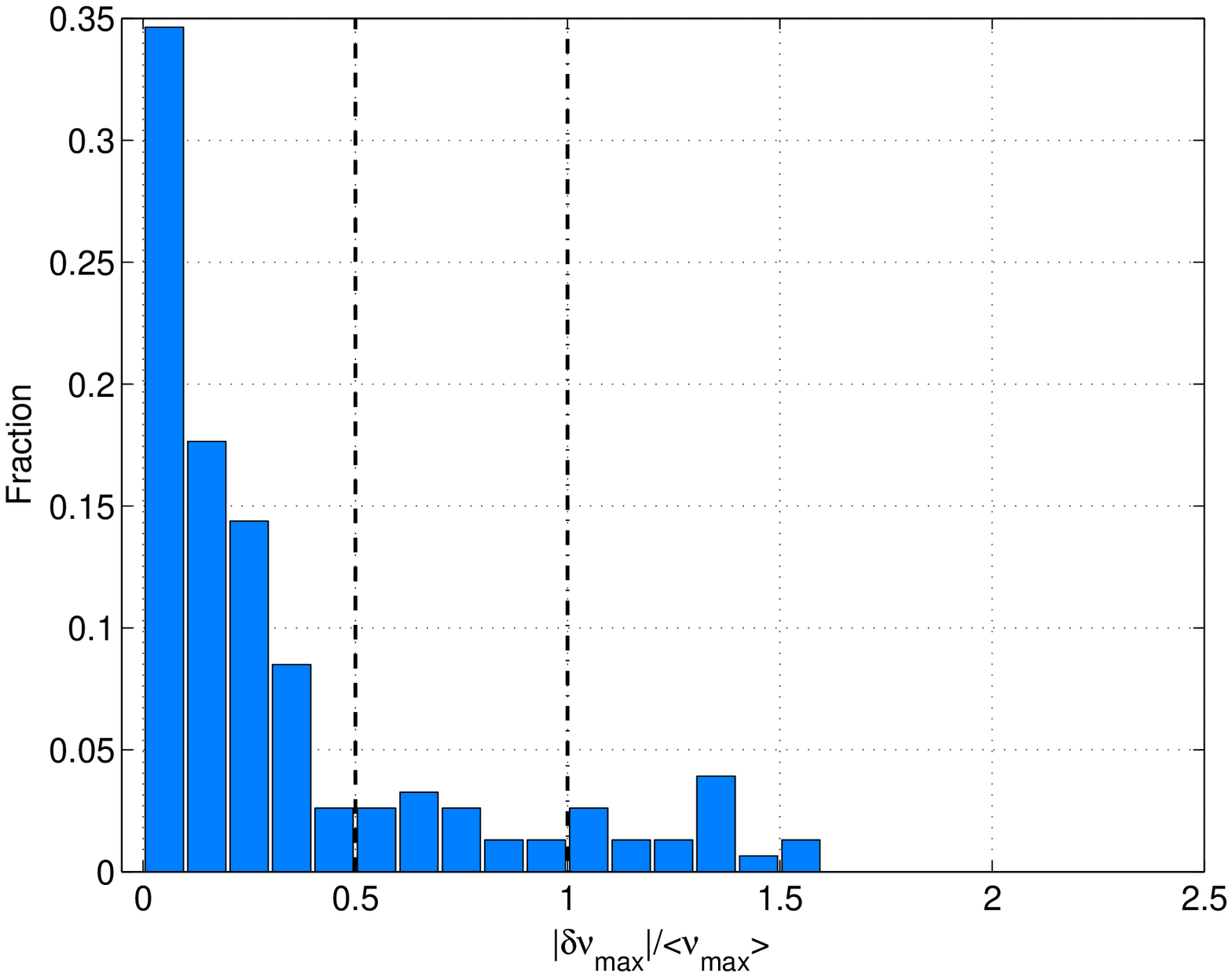}

\caption{Upper-left panel: Mass-ratio distribution of binaries in a
  TRILEGAL simulation (green bars) based on a uniform IMRD with
  $f_{\rm b}=0.3$ and $b_{\rm b}=0.7$ (see Section~\ref{sec:tri}). The
  MRD of detectable seismic binaries is illustrated by wide yellow
  bars, {while narrow orange bars show the IMRD}.  {Upper-right panel:
    mass distribution of primary components in asteroseismic binaries
    (narrow magenta bars), of primary components in binaries in which
    oscillations can be detected in only one component (dark blue
    bars), and of single stars with detectable oscillations (wide
    light-blue bars).}  Bottom panel: Histogram of the difference in
  the $\nu_{\rm max}$ of the components of seismically detected
  binaries (from the TRILEGAL simulations), normalized by the average
  of the two $\nu_{\rm max}$ (see text).}

 \label{fig:binpars} 
 \end{figure}

Our simulations indicate that the mass ratios of the detectable
binaries will lie close to unity (within 10\,\%; see upper-left panel
of Fig.~\ref{fig:binpars}). We find this result to be largely
independent of the assumed IMRD. Many of the detectable binaries will
have He-core-burning stars, where the observed mass ratios may have
been affected by mass loss and hence altered from the initial
values. However, in the simulations -- which use the prescription for
RGB mass loss of \citet{Reimers75} -- we find that the observed MRD of
asteroseismic binaries is an excellent proxy for the initial MRD
(Fig.~\ref{fig:binpars}, upper-left panel).

The mass distribution of primary components in asteroseismic binaries
is peaked at a higher value compared to single stars, or primary
components in binaries in which oscillations can be detected in only
one component (Fig.~\ref{fig:binpars}, upper-right panel). We ascribe
this to the longer duration (relative to the main-sequence) of the
core-He burning phase in higher-mass giants.

Most detectable seismic binaries will have oscillation spectra that
overlap considerably in frequency. The observed power in the
oscillations spectrum of a solar-type or red-giant star is modulated
in frequency by an envelope that typically has an approximately
Gaussian shape. With $\nu_{\rm max}$ denoting the frequency at which
the oscillations present their strongest observed power, the
\textsc{fwhm} of the envelope is to good approximation equal to
$\nu_{\rm max}/2$ \citep{Mosser2010, Chaplin11b}. The bottom panel of Fig.~\ref{fig:binpars} plots
the difference in the $\nu_{\rm max}$ of the components of
asteroseismic binaries (from the TRILEGAL simulations), normalized by
the average of the two $\nu_{\rm max}$.  When that fractional
difference lies within $\pm 0.5$, the peak of the oscillations
spectrum of each star lies within the \textsc{fwhm} of its companion;
the spectra therefore overlap significantly. Only when the normalized
difference exceeds unity is the observable power of the two
oscillations spectra all but separated in frequency. The bottom panel
of Fig.~\ref{fig:binpars} shows that cases where the spectra are well
separated should be rare.

\begin{figure}
 \includegraphics[angle=0, width=.5\textwidth]{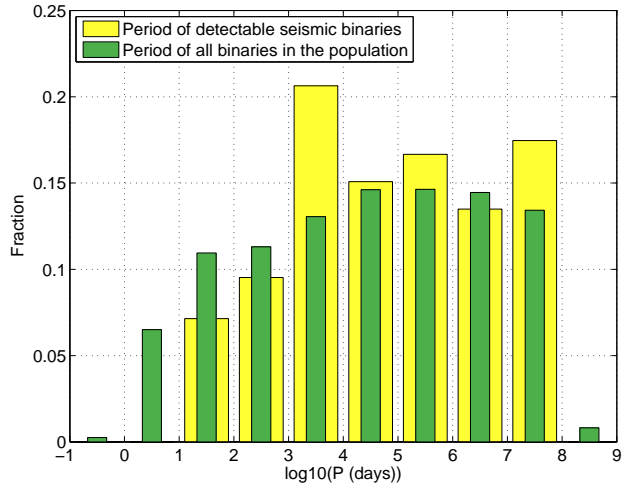}

\caption{ Orbital period distribution of all binaries (narrow green
  bars) and detectable seismic binaries (wide yellow bars) in a BiSEPS
  simulation computed assuming $s=0$ (see Section~\ref{sec:biseps}).}

 \label{fig:period} 
 \end{figure}

The distribution of the orbital periods shown by the artificial
asteroseismic binaries is plotted in Fig.~\ref{fig:period}. In general
it follows the distribution shown by all binaries, but displays a
short-period cut-off near 40\,days, reflecting the need to accommodate
two giant stars.

 \section{{Conclusions}}
 \label{sec:disc}

We have used results from two stellar population simulations of the
\emph{Kepler} field of view to consider {potential detection
  yields} from applying a novel asteroseismic method to discover
binary systems comprised of solar-type stars and/or red giants. The
method relies on the detection of the presence of two solar-like
oscillation spectra in the frequency spectrum of a single
\emph{Kepler} lightcurve. Our predictions suggest that 200 or more
binaries may be detectable in this manner using the \emph{Kepler}
data. Most of these binaries should be comprised of two
He-core-burning red giants. Owing largely to the limited numbers of
targets with the requisite short-cadence \emph{Kepler} data, we expect
only a small number of detected binaries containing solar-type stars.

The method is not biased by, or dependent on, the inclination, period
(separation) or velocities of the constituent components. As such, it
may provide the only way to detect some of the binaries that are in
the \emph{Kepler} database. The predicted yield of detections is
sensitive to the IMRD, which is a robust indicator of the pristine
mass distribution at the epoch of star formation (e.g., see
\citealt{parker13}). Given the additional constraints, the detection
of asteroseismic binaries will also provide targets that are useful
for testing stellar models and will add significantly to the list of
eclipsing binaries with a solar-like oscillating component
\citep[e.g., see][]{Hekker10, Gaulme13}.

{Work is now underway to discover asteroseismic binaries in the
  \emph{Kepler} data archive.}

\acknowledgments

The authors acknowledge useful discussions with J.~Ballot,
R.~Garc{\'i}a, S.~Hekker, D.~Stello and T.~White.  W.J.C., Y.E., A.M.,
R.F. and U.K. acknowledge support from the UK Science and Technology
Facilities Council (STFC). Funding for the Stellar Astrophysics Centre
is provided by The Danish National Research Foundation (Grant
agreement no.: DNRF106).  The research leading to these results has
received funding from the European Community's Seventh Framework
Programme (FP7/2007-2013) under grant agreement no. 312844
(SPACEINN). We are grateful for support from the International Space
Science Institute (ISSI). {Finally, we also thank the anonymous
  referee for comments that helped to improve the paper.}

\clearpage


%


\end{document}